\documentclass[preprint, preprintnumbers, amsmath ,amssymb, letterpaper,  superscriptaddress, showpacs,  prl]{revtex4}

\usepackage{graphicx}
\usepackage{dcolumn}
\usepackage{bm}

\usepackage{subfigure}

\preprint{}

\begin{document}

\title
{
Generalized Kapchinskij-Vladimirskij Distribution and Beam Matrix for Phase-Space Manipulations of High-Intensity Beams
}
\author{Moses Chung} \email{mchung@unist.ac.kr}
\affiliation{Department of Physics, Ulsan National Institute of Science and Technology, Ulsan 689-798, Korea}
\author{Hong Qin}
\affiliation{Plasma Physics Laboratory, Princeton University, Princeton, New Jersey 08543}
\affiliation{Department of Modern Physics, University of Science and Technology of China, Hefei, Anhui 230026, China}
\author{Ronald C. Davidson} \thanks{Deceased.}
\affiliation{Plasma Physics Laboratory, Princeton University, Princeton, New Jersey 08543}
\author{Lars Groening}
\affiliation{GSI Helmholtzzentrum f$\ddot{u}$r Schwerionenforschung GmbH, Planckstrasse 1, D-64291 Darmstadt, Germany}
\author{Chen Xiao}
\affiliation{GSI Helmholtzzentrum f$\ddot{u}$r Schwerionenforschung GmbH, Planckstrasse 1, D-64291 Darmstadt, Germany}
\date{\today}

\begin{abstract}
In an uncoupled linear lattice system,
the  Kapchinskij-Vladimirskij (KV) distribution, formulated on the basis of the single-particle Courant-Snyder (CS) invariants,
has served as a fundamental theoretical basis for the analyses of the equilibrium, stability, and transport properties of high-intensity beams for the past several decades.
Recent applications of high-intensity beams, however, require beam phase-space manipulations by intentionally introducing strong coupling.
In this Letter,
we report the full generalization of the KV model by including
all of the linear (both external and space-charge) coupling forces,
beam energy variations, and arbitrary emittance partition, which all form
essential elements for phase-space manipulations.
The new generalized KV model yields spatially uniform density profiles and corresponding linear self-field forces as desired.
The corresponding matrix envelope equations and beam matrix for the generalized KV model
provide important new theoretical tools for the detailed design and analysis of high-intensity beam manipulations,
for which previous theoretical models are not easily applicable.
\end{abstract}

\maketitle

For the past several decades,
the well-known Courant-Snyder (CS) theory \cite{CS} has served as a fundamental theoretical tool
in designing and analyzing an {\it uncoupled} linear lattice system.
One of the recent areas of investigation by the beam physics community, however,
is to manipulate the beam phase-space by intentionally introducing strong coupling.
The round-to-flat beam transformation \cite{Kim2003, Sun2004,RTB1,RTB2,RTB3} and
the transverse-to-longitudinal emittance exchange \cite{Emma2002, Ruan2011, EE1, EE2}
have been investigated for electron injectors.
The generation of flat hadron beams has recently drawn significant attention
in the context of optimizing emittance budgets in heavy ion synchrotrons \cite{Lars2013, Lars2014},
and improving space-charge and beam-beam luminosity limitations in colliders \cite{Burov}.
For muon ionization cooling,
special arrangements of solenoidal mangets are employed to achieve six-dimensional emittance reduction \cite{HCC, Diktys}.

Various attempts have been made to extend the uncoupled CS theory to the case of general linear coupled systems \cite{Lee-Teng1, Mais-Ripken1, Lebedev, Wolski.book}.
However, due to the lack of a proper CS invariant for the coupled  dynamics,
previous analyses did not retain the elegant mathematical structure present in the original CS theory.
Recently, Qin et al. \cite{GCS_PRL, GCS.PRSTAB} have identified the generalized CS invariant for
the linear coupled systems including
both solenoidal and skew-quadrupole magnets, and variation of beam energy along the reference orbit.
For phase-space manipulations,
solenoidal and  skew-quadrupole magnets are frequently used to provide strong coupling, as mentioned previously.
Moreover, the relativistic mass increase may be important when there is a rapid acceleration of low-energy  beams.

For some of the beam manipulations, space-charge effects are non-negligible as well;
hence in those cases, we require a further generalization that incorporates  space-charge effects into linear coupling lattices.
In the original CS theory, space-charge effects were considered by means of the Kapchinskij-Vladimirskij (KV) distribution \cite{KV}.
For an intense beam propagating trough an alternating-gradient lattice,
the KV distribution is the only known exact solution to the nonlinear Vlasov-Maxwell equations \cite{Davidson.book, LundUSPAS},
and it generates linear space-charge forces consistent with the CS theory.
Through the concept of rms-equivalent beams \cite{Sacherer_RMS, Reiser.book, LundUSPAS},
the KV beam model remains the most important basic design tool for high-intensity beam transport,
even in the presence of nonlinear space-charge contributions.
Several generalizations have been proposed for the KV model in order that it can be applied to coupled systems as well
\cite{Sacherer, Chernin, Z.Phys.C, Banard_PAC, G_KV, G_KV2}.
However, none of them  incorporates the solenoids and skew-quadrupoles simultaneously
with a proper CS invariant.

In this Letter, we report the first complete generalization of the KV model for the general linear coupled system,
so that the model describes all of the important processes for transverse phase-space manipulations of high-intensity beams.
Due to the existence of the generalized CS invariant,
the KV model developed here provides a self-consistent solution to the nonlinear
Vlasov-Maxwell equations for high-intensity beams
in coupled lattices,
and leads to a matrix version of the envelope equation with an elegant Hamiltonian structure.
We emphasize that space-charge effects during emittance manipulation,
illustrated by a numerical example in this Letter, is one area that previous KV models could not address.


First, we consider a transverse Hamiltonian in general linear focusing lattice
of the form
\begin{equation}
H_\perp = \frac{1}{2} {\bf z}^T A_c(s) {\bf z}, ~~~
A_c(s) = \left[
        \begin{array}{cc}
         \kappa & R \\
          R^T   & m^{-1} \\
        \end{array}
      \right].
      \label{Hamiltonian}
\end{equation}
Here, ${\bf z} = (x, y, p_x, p_y)^T$ denotes the transverse canonical coordinates, $s$ is the path length that plays the role of a time-like variable,
and $\kappa$ and $m^{-1}$ are $2\times2$ symmetric matrices.
The arbitrary $2\times2$ matrix $R$ is not symmetric in general.
The canonical momenta are normalized by a fixed reference momentum $p_0 = \gamma_0 m_b \beta_0 c$.
Based on the generalized CS theory developed in Refs. \cite{GCS_PRL, GCS.PRSTAB},
we obtain the solution for the coupled dynamics governed by the Hamiltonian (\ref{Hamiltonian}) in the form of a linear map
$ {\bf z}(s) = M(s) {\bf z}_0 $,
where ${\bf z}_0$ is the initial condition and $M$ is the transfer matrix defined by
\begin{equation}
M(s)
=
Q^{-1} P^{-1} P_0 Q_0
= \left[
      \begin{array}{cc}
        W & 0 \\
        V & W^{-T} \\
      \end{array}
    \right]
    P^T
    \left[
      \begin{array}{cc}
        W^{-1} & 0 \\
        - V^T & W^T \\
      \end{array}
    \right]_0,
    \label{M_compact}
\end{equation}
where subscript ``0'' denotes initial conditions at $s=0$,
$P^T = P^{-1}$ and $P$ is a symplectic rotation,
and $P_0$ is set equal to the unit matrix $I$ without loss of generality.
Here, the $2\times2$ matrices $W$ and $V$ are defined by
$
W = w^T
$ and
$
V =  m \left( \frac{dw^{T}}{ds}  - R^T w^T \right)
$.
Furthermore, the $2 \times 2$ envelope matrix $w$ is obtained by solving the matrix envelope equation given by \cite{GCS_PRL,GCS.PRSTAB}
\begin{equation}
\frac{d}{d s} \left( \frac{dw}{ds} m - w Rm  \right) + \frac{dw}{ds} m R^T + w ( \kappa -R m R^T) - \left( w^T w m w^T \right)^{-1} = 0.
\label{env}
\end{equation}
We note that the second-order matrix differential equation (\ref{env}) can be expressed
in terms of two first-order equations, i.e.,
\begin{equation}
W' = m^{-1} V + R^T W, ~~~
V' = - \kappa  W - R V + \left( W^T m W W^T \right)^{-1}. \label{V'}
\end{equation}
The variable $V$ can be considered to be the matrix associated with the envelope momentum \cite{Accelerator_physics3}.
We also note that Eq. (\ref{V'}) has similar Hamiltonian structure   to the single particle equations of motion
except for  the term $\left( W^T m W W^T \right)^{-1}$ [see Eq. (\ref{xp2}) for comparison and Ref. \cite{Moses.PoP4} for a more detailed discussion].

The $4\times4$ phase advance matrix $P$ has the following form \cite{GCS_PRL,GCS.PRSTAB}
\begin{equation}
P = \left[
      \begin{array}{rr}
        C_o & - S_i \\
        S_i &   C_o \\
      \end{array}
    \right].
\end{equation}
Here, $C_o$ and $S_i$ are the $2\times2$ matrices that satisfy
$
C_o' = - S_i \left( W^T m W \right)^{-1}
$
and
$
S_i' = + C_o \left( W^T m W \right)^{-1}
$,
where the term $\left( W^T m W \right)^{-1}$ represents the phase advance rate.
From the symplecticity of $P$, we  note that
$S_i C_o^T = C_o  S_i^T$ and $S_i S_i^T + C_o C_o^T = I$.
The generalized CS invariant of the Hamiltonian (\ref{Hamiltonian}) is given by
$
I_\xi = {\bf z}^T Q^T P^T \xi P Q {\bf z},
$
where $\xi$ is a constant $4\times4$ matrix, which is both symmetric and positive definite.
The $\xi$ matrix acquires a meaning associated with emittance  when the beam distribution is defined in terms of the CS invariant $I_\xi$
[see, for example, Eq. (\ref{delta})].
The two symplectic eigenvalues of $\xi$ are directly connected to the eigen-emittances of the beam  \cite{Moses.PoP2}.

By using $s$ as an independent coordinate,
and  treating $ |p_x - q_b A_x / p_0|, |p_y - q_b A_y / p_0| \ll p_0$ and  $|q_b \phi^{sc}| \ll \gamma_b m_b c^2 $,
we can express the transverse Hamiltonian (normalized by $p_0$) to second order in the transverse momenta as \cite{Davidson.book, Wolski.book}
\begin{equation}
H_\perp =
\frac{1}{2 p_b / p_0} \left[ \left(p_x - \frac{q_b A_x}{p_0} \right)^2 + \left(p_y - \frac{q_b A_y}{p_0}\right)^2   \right]
+ \left(  \frac{1}{\gamma_b^2 } \right) \frac{q_b \phi^{sc}}{\beta_b c p_0}
- \frac{q_b A_s^{ext}}{p_0},
\end{equation}
where we have used the fact that
the longitudinal vector potential is composed of both external ($A_s^{ext}$) and self-field ($A_s^{sc}$) contributions,
and the self-field potentials $\phi^{sc}$ and $A_s^{sc}$ are related approximately by $A_s^{sc} = \beta_b \phi^{sc} /c$.
Also, it is assumed that the reference trajectory is a straight line, that the longitudinal motion is independent of the transverse motion,
and that there is no external electric focusing.
Furthermore, $p_b(s) = \gamma_b m_b \beta_b c$,  $\gamma_b(s)$, and  $\beta_b(s)$ are regarded as prescribed functions of $s$ set by the
acceleration schedule of the beamline \cite{LundUSPAS}.
Hence, for a combination of the quadrupole, skew-quadrupole, and solenoidal fields, we obtain
the following matrices for the Hamiltonian (\ref{Hamiltonian})
\begin{equation}
\kappa_{ext}(s) = \left[
              \begin{array}{cc}
                \kappa_q + \left( \frac{ \beta_0 \gamma_0}{ \beta_b \gamma_b} \right) \Omega_L^2 & \kappa_{sq} \\
                \kappa_{sq} & -\kappa_q + \left( \frac{ \beta_0 \gamma_0}{ \beta_b \gamma_b} \right) \Omega_L^2 \\
              \end{array}
            \right],
\end{equation}
\begin{equation}
R(s) = \left[
         \begin{array}{cc}
           0 & - \left( \frac{ \beta_0 \gamma_0}{ \beta_b \gamma_b} \right)\Omega_L \\
           \left( \frac{  \beta_0 \gamma_0}{ \beta_b \gamma_b} \right) \Omega_L & 0 \\
         \end{array}
       \right], ~~~
m^{-1}(s) = \left[
         \begin{array}{cc}
           \left( \frac{ \beta_0 \gamma_0}{ \beta_b \gamma_b} \right) & 0 \\
           0 & \left( \frac{ \beta_0 \gamma_0}{  \beta_b \gamma_b} \right) \\
         \end{array}
       \right].
\end{equation}
Here, $\kappa_q = q_b B'_{q}(s) / p_0$, $\kappa_{sq} = q_b B'_{sq}(s) / p_0$, and $\Omega_L = q_b B_s (s) / 2 p_0$.

For low energy (i.e., $\beta_b \ll 1$) beams, we note that
the longitudinal acceleration acts to damp particle oscillations more rapidly \cite{LundUSPAS}.
In such cases, so-called reduced coordinates are often
introduced to avoid the complication due to the acceleration \cite{Chun-xi}.
Since the new generalized KV model has been formulated in terms of the canonical momenta
with the relativistic mass increase already included,
an additional transformation to the reduced coordinates is unnecessary.

Since the focusing matrix $\kappa$ used in the generalized CS theory is an arbitrary $2\times2$ symmetric matrix,
we can include the coupled linear space-charge force as
\begin{equation}
- \kappa {\bf x} = - \kappa_{ext} {\bf x} -  \kappa_{sc} {\bf x} = - \kappa_{ext} {\bf x} -  \left( \frac{ \beta_0 \gamma_0^2}{ \beta_b \gamma_b^2} \right)  \nabla \psi,
\label{kappa_total}
\end{equation}
where ${\bf x} = (x, y)^T$, and $\kappa_{ext}$ is constructed from the external lattices.
The normalized self-field potential is  defined by
$\psi = q_b \phi^{sc} / \gamma_0^2 \beta_0 c p_0 $.
In this coupled linear focusing system,
$\psi ({\bf x}, s)$ and the beam distribution function $f ({\bf x}, {\bf p}, s)$ evolve according to
\begin{equation}
  \frac{\partial f}{\partial s} + {\bf x'} \cdot \frac{\partial f}{\partial {\bf x}}
+ \left[-\kappa_{ext} {\bf x} -  \left( \frac{ \beta_0 \gamma_0^2}{ \beta_b \gamma_b^2} \right) \nabla \psi - R {\bf p} \right] \cdot \frac{\partial f}{\partial {\bf p}} = 0 ,
\label{Vlasov}
\end{equation}
\begin{equation}
\nabla^2 \psi = -\frac{2 \pi K_b}{N_b} \int f dp_x dp_y =  -\frac{2 \pi K_b}{N_b} n.
\label{Maxwell}
\end{equation}
Here, ${\bf x'} = (x', y')^T$ is the normalized transverse velocity,
and ${\bf p} = (p_x, p_y)^T$ is the normalized canonical momentum defined from the Hamiltonian equations of motion
($d {\bf z}/ ds = J A_c {\bf z}$, where $J$ is the unit symplectic matrix) as
\begin{equation}
{\bf x'} = m^{-1}{\bf p} + R^T {\bf x}, ~~~
{\bf p}' = - \kappa {\bf x} - R {\bf p} .
\label{xp2}
\end{equation}
The self-field perveance is defined by $K_b = (1/ 4 \pi \epsilon_0) (2 N_b q_b^2 / \gamma_0^2 \beta_0 c p_0 )$ in SI units,
and the line density $N_b = \int f dx dy dp_x dp_y$ is assumed to be constant.
Based on the analysis in Ref. \cite{G_KV2}, we consider the following distribution function
\begin{equation}
f = \frac{N_b \sqrt{| \xi |}}{\pi^2} \delta ( I_\xi - 1).
\label{delta}
\end{equation}
which is a solution of the Vlasov equation
(i.e., $ d f / d s = 0 $ because $I_\xi$ is a constant of motion),
and generates the coupled linear space-charge force (i.e., $\int f dp_x dp_y$ is spatially uniform in the beam interior).

Using the Cholesky decomposition method,
the momentum integral in Eq. (\ref{Maxwell}) can be carried out in a straightforward manner.
First, we decompose $Q^T P^T \xi P Q$ in terms of a lower triangular matrix $L$ according to
$Q^T P^T \xi P Q = L^T L$,
and then introduce new coordinates ${\bf Z} = (X, Y, P_X, P_Y)^T= L {\bf z}$ defined by
$
{\bf X} =  ( X, Y )^T =   \bar{D}^{T/2} W^{-1} {\bf x}
$ and
$
{\bf P} =  ( P_X, P_Y )^T =  ( D^{-1/2} B^T W^{-1} - D^{T/2} V^T ) {\bf x} + D^{T/2} W^T {\bf p}
$.
We note that, similar to the original KV model, the distribution function $f$ in Eq. (\ref{delta})
represents the trajectories of all particles lying on the surface of the 4D hyper-ellipsoid, ${\bf Z}^T {\bf Z} = X^2 + Y^2 + P_X^2 + P_Y^2 = 1$ \cite{Reiser.book}.
Here, the square-root of a symmetric and positive definite matrix $D$
is defined by $D^{1/2} D^{T/2} = D$.
The $\bar{D}$ matrix is known as the Schur complement of $D$, and it has the following definitions and properties.
\begin{equation}
P^T \xi P = \left[
              \begin{array}{cc}
                A   & B \\
                B^T & D \\
              \end{array}
            \right]
          = \left[
              \begin{array}{cc}
                \bar{D}^{1/2} & B D^{-T/2} \\
                0 &  D^{1/2} \\
              \end{array}
            \right]
            \left[
              \begin{array}{cc}
                \bar{D}^{T/2} & 0 \\
                 D^{-1/2} B^T   & D^{T/2} \\
              \end{array}
            \right],
\end{equation}
where $\bar{D} = A - B D^{-1} B^T = \bar{D}^T$  and $ |P^T \xi P| = |\xi| = |\bar{D}| |D|$.

The Jacobians of the linear coordinate transformations are given by
$dX dY = |\bar{D}^{T/2} W^{-1}| d x dy$, and $ dP_X dP_Y = |D^{T/2} W^T| d p_x d p_y$.
Then, it can be readily shown that the number density $n(x,y,s)$ of the beam particles is given by
\begin{equation}
n(x, y, s) = \int  f dp_x dp_y = \left\{
               \begin{array}{ll}
                 N_b \frac{ | \bar{D}^{T/2} W^{-1} | }{\pi}, &   0 \le {\bf X}^T {\bf X} < 1, \\
                 0,                        &   1 < {\bf X}^T {\bf X},
               \end{array}
             \right.
             \label{density}
\end{equation}
where $\int n (x,y,s) dx dy =N_b$ is the line density.
From Eq. (\ref{density}), we note that $n$ is spatially uniform and a function only of  $s$.
The boundary of the beam is determined from
${\bf X}^T {\bf X} = {\bf x}^T ( W^{-T} \bar{D} W^{-1} ) {\bf x} = 1$,
which is a tilted ellipse in $(x,y)$ space with area equal to $\pi | \bar{D}^{T/2} W^{-1} |^{-1}$.
The transverse dimensions of the tilted ellipse,  $a$ and $b$, are
determined by the two eigenvalues $(1/a^2, 1/b^2)$  of the matrix $W^{-T} \bar{D} W^{-1}$.
Therefore, the coupled linear space-charge force coefficient $\kappa_{sc}$
can be expressed as
\begin{equation}
\kappa_{sc} = - \left( \frac{ \beta_0 \gamma_0^2}{ \beta_b \gamma_b^2} \right) \frac{2 K_b}{a + b}
G \left[
    \begin{array}{cc}
      1/a & 0   \\
      0   & 1/b \\
    \end{array}
  \right]
G^{-1}.
\label{kappa_sc}
\end{equation}
Here, $G$ is the matrix  constructed by the two normalized eigenvectors ${\bf v}_1$
and ${\bf v}_2$ of the matrix $W^{-T} \bar{D} W^{-1}$ as
$G = ({\bf v}_1, {\bf v}_2)$.
Note that $G$ is a rotation matrix, i.e., $G^{-1} = G^T$.
When the space-charge force term $\kappa_{sc}$ is substituted back into Eq. (\ref{kappa_total}),
the envelope equations (\ref{V'}) become a set of closed nonlinear matrix equations
for the envelope matrix $W$ and its associated envelope momentum matrix $V$.

To demonstrate the exact connection between $Q^T P^T \xi P Q$ and the beam matrix,
we introduce the geometric factor $g$ \cite{Davidson.book} and the symmetric matrix $\Sigma$ defined by
$ Q^T P^T \xi P Q = g \Sigma^{-1} $.
We will show that there exits a real number $g$ which makes $\Sigma$ equal to the beam matrix $\left<{\bf z}{\bf z}^T\right> $,
in which $\left< \cdots \right>$ denotes statistical average over the distribution function $f$.
Since the  matrix $\Sigma$ is  real and symmetric,
we consider the eigenvalue equation for $\Sigma$ given by $\Sigma {\bf u}_i  = \lambda_i {\bf u}_i$.
We  can then make use of the orthonormality of the eigenvectors \cite{Bishop.book}
to express  ${\bf z} = \sum_{j = 1}^4 y_j {\bf u}_j$, where $y_j = {\bf u}_j^T {\bf z}$.
It then follows that
\begin{eqnarray}
\left< {\bf z}{\bf z}^T \right>
= \frac{\sqrt{ |\xi| }}{\pi^2}  \sum_{i = 1}^4 {\bf u}_i {\bf u}_i^T
\int  \delta ( g \sum_{k=1}^4 \frac{ y_k^2}{ \lambda_k} - 1 ) y_i^2 d {\bf y}.
\end{eqnarray}
Here, we have used the fact that the above integral  vanishes by symmetry unless $y_i = y_j$.
After some straightforward algebra, the above integral yields
$
(\lambda_i / g) \left[ \prod_{j=1}^4 \sqrt{\lambda_j / g} \right] ( \pi^2 / 4)
$.
We then finally obtain
$
\left< {\bf z}{\bf z}^T \right>
= \frac{1}{4 g}  \sum_{i = 1}^4 {\bf u}_i {\bf u}_i^T   \lambda_i  = \frac{1}{4 g} \Sigma
$.
Therefore, if $g = 1/4$,  then $\Sigma = \left< {\bf z}{\bf z}^T \right> = \frac{1}{4} Q^{-1} P^{-1} \varepsilon P^{-T} Q^{-T}$,
where the emittance matrix is defined by $\varepsilon = \xi^{-1}$.
We note that the transverse rms emittance is
$\epsilon_\perp^2 = \sqrt{|\Sigma|} = \frac{1}{16} \sqrt{|\varepsilon|}$.
This is the natural generalization of the original KV model,
in which the total (or 100\%) emittance is 4 times larger than the rms emittance for each transverse phase-space.

Once the initial beam matrix $\Sigma_0$ is prescribed,
the beam matrix at an arbitrary position $s$ can be calculated in terms of the transfer matrix $M$ as
$ \Sigma(s) = M \Sigma_0 M^T$.
In principle, the transfer matrix $M$ is independent of the choice of the parametrization because $M$ is solely determined by the equations of motion.
Therefore, the envelope equations (\ref{V'}) can be solved for arbitrary choices of the initial conditions $(W, V)_0$.
Furthermore, for the case of negligible space-charge,
the envelope equations (\ref{V'}) become independent of the initial beam matrix $\Sigma_0 = \frac{1}{4} Q_0^{-1} \varepsilon Q_0^{-T} $ as well.
On the other hand, for the case of intense space-charge,
the beam envelopes evolve under the influence of the beam matrix $\Sigma$,
because the space-charge focusing coefficient $\kappa_{sc}$ depends on $\Sigma$.
Hence, in this case, it is important to ensure that the generalized CS parametrization generates the initial beam matrix $\Sigma_0$ correctly.
This can be achieved by requiring
$ \varepsilon = \xi^{-1} =  4 Q_0 \Sigma_0 Q^T_0 $.
When $Q_0$ is calculated for specified initial conditions $(W, V)_0$,
the ten free parameters in $\varepsilon$ (or $\xi^{-1}$) are determined accordingly.
Note that when different initial conditions $(W, V)_0$ are used,
the emittance matrix $\varepsilon$ itself is calculated differently; however, it generates the same $\Sigma_0$, $M$, and  eigen-emittances.

Based on Refs. \cite{Kim2003, Lars2013},
we specify the initial beam matrix of a cylindrically symmetric beam in the following form
\begin{equation}
\Sigma_0 =
\left[
  \begin{array}{cccc}
    \sigma^2            & 0                     & 0                                      & \kappa_0 \sigma^2 \\
    0                   &   \sigma^2            & - \kappa_0 \sigma^2                    & 0 \\
    0                   & - \kappa_0 \sigma^2   &   \sigma'^2 + \kappa_0^2 \sigma^2      & 0 \\
    \kappa_0 \sigma^2   & 0                     & 0                                      & \sigma'^2 + \kappa_0^2 \sigma^2 \\
  \end{array}
\right],
\label{Sigma0}
\end{equation}
where $\sigma^2 = \left< x^2 \right> = \left< y^2 \right>$, and $\sigma'^2 = \left< x'^2 \right> = \left< y'^2 \right>$.
It can be shown that the two eigen-emittances are given by
$
\epsilon_{1,2} = \epsilon_{\rm eff} \pm \mathcal{L}
$, where
$\mathcal{L} = \kappa_0 \sigma^2$ and
$\epsilon_{\rm eff} = \sqrt{ (\sigma \sigma')^2 + \mathcal{L}^2}$.
The initial beam matrix $\Sigma_0$ in the form of Eq. (\ref{Sigma0}) can be obtained,
either by generating an electron beam inside a solenoid as in the round-to-flat beam (RTFB) transformation experiment \cite{Sun2004},
or by stripping an ion beam inside a solenoid as in the emittance transfer experiment (EMTEX) \cite{Lars2013,Lars2014}.
For the RTFB transformation experiment, $\kappa_0$ is given by $\kappa_0 = \frac{B_s}{2 (B\rho)_c}$,
where $B_s$ and $(B \rho)_c$ are the solenoidal magnetic field and beam rigidity at the cathode, respectively.
For the EMTEX experiment, $\kappa_0 = \left[ \frac{(B\rho)_{\rm in}}{(B\rho)_{\rm out}} - 1\right] \frac{B_s}{2 (B\rho)_{\rm in}}$,
where $(B \rho)_{\rm in}$ and $(B \rho)_{\rm out}$ are the beam rigidity before and after the stripping foil, respectively.
To remove the correlation in $\Sigma_0$,
a beam-line constructed by three skew-quadrupoles is often used \cite{Kim2003, Lars2013}.

\begin{figure}[tb]
  \centering
  \includegraphics[width= 12cm]{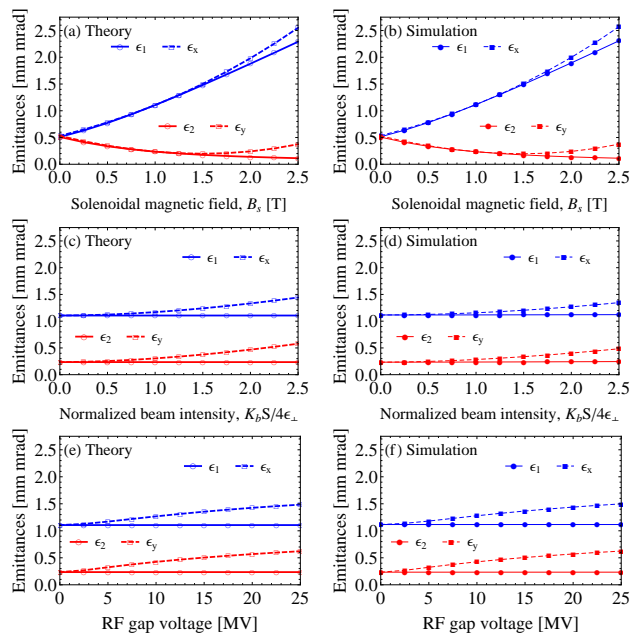}\\
  \caption{
  Plots of the eigen-emittances (solid lines with circles) and projected rms emittances (dashed lines with squares)
  at the exit of the skew-quadrupole triplet.
  Frames (a), (c), and (e) represent the results of the present KV model analyzed by MATHEMATICA  \cite{Mathematica}, and
  frames (b), (d), and (f) represent the results of the multi-particle tracking simulations using TRACK code \cite {TRACK}.
  Frames (a) and (b) correspond to the cases with $K_b = 0$ and zero acceleration;
  frames (c) and (d) to the cases with $B_s = 1$ T and zero acceleration;
  frames (e) and (f) to the cases with $B_s = 1$ T and $K_b = 0$,   respectively.
  }
  \label{KV_Fig1}
\end{figure}

As a numerical example,
we consider an initial beam matrix with parameters of the EMTEX  in Ref. \cite{Lars2013}.
The focusing coefficients of the skew-quadrupoles are kept fixed at
the values used to decouple the beam produced by a solenoidal field of 1.0 T,
with the conditions of zero space-charge and zero acceleration.
Figures \ref{KV_Fig1}(a) and \ref{KV_Fig1}(b) indicate that
the decoupling processes are not sensitive to the solenoidal field strength $B_s$,
particulary when $B_s \lesssim 1.5$.
This tendency has been investigated in detail in Refs. \cite{Lars2014, Lars_arXiv}.
Therefore, for a given skew-quadrupole triplet setting, one can obtain arbitrary emittance ratios
by simply changing the single parameter $B_s$.
Figures \ref{KV_Fig1}(c) and \ref{KV_Fig1}(d) show the
effects of the space-charge forces on the decoupling processes.
If the normalized beam intensity $K_b S / 4 \epsilon_\perp$ (in which $S$ is the axial periodicity length or the characteristic length of the beamline)
is greater than about 1.0 (i.e., the space-charge force becomes comparable to or greater than the emittance contribution),
the deviations of the rms emittances from the eigen-emittances become significant and increase continuously.
Conventional multi-particle tracking simulations including space-charge effects show
a good agreement ($< 7\%$ of relative errors in projected rms eimttances) with the present KV model.
Figures \ref{KV_Fig1}(e) and \ref{KV_Fig1}(f) show the effects of the beam energy variation on the decoupling processes.
The rms emittances deviate from the eigen-emittances  when an RF voltage is applied to the acceleration gap located
between the solenoid and the skew-quadrupole triplet.

In summary,
we have fully generalized the KV model by including all the linear coupling elements,
so that it provides a new advanced theoretical tool for the design and analysis of complex beamlines with strong coupling.
In the numerical example summarized in Fig. \ref{KV_Fig1},
we have demonstrated the usefulness and effectiveness of the new generalized KV model
in understanding phase-space manipulations of high-intensity beams,
for which previous KV models are  inapplicable.

\begin{acknowledgments}
This research was supported
by  the National Research Foundation of Korea (NRF-2015R1D1A1A01061074).
This work was also supported by the U.S. Department of Energy Grant No. DE-AC02-09CH11466.
\end{acknowledgments}



\end{document}